\documentstyle[11pt]{article}
\begin{document}
\date{TRI-PP-94-4\\January 1994}
\title{The off--shell electromagnetic form factors of pions and kaons
in chiral perturbation theory}
\author{T.\ E.\ Rudy, H.\ W.\ Fearing and S.\ Scherer\\
TRIUMF, Vancouver, British Columbia, Canada V6T 2A3}
\maketitle
\begin{abstract}
The off--shell electromagnetic vertex of a (pseudo--) scalar particle
contains, in general, two form factors $F$ and $G$ which depend, in addition
to the squared momentum transfer, on the invariant masses associated with the
initial and final legs of the vertex.
Chiral perturbation theory to one loop is used to calculate the off--shell
form factors of pions and kaons.
The formalism of Gasser and Leutwyler, which was previously used to calculate
the on--shell limit of the form factor $F$, is extended to accommodate the most
general form for off--shell Green's functions in the pseudoscalar meson sector.
We find that chiral symmetry predicts that the form factors $F$ of the
charged pions and kaons go off--shell in the same way, i.e., the
off--shell slope at the real photon point is given by the same new
phenomenological constant $\beta_1$.
Furthermore, it is shown that at order $p^4$ the form factor $F$ of
the $K^0$ does not show any off--shell dependence.
The form factors $G$ are all related to the form factors $F$ in the
correct fashion as required by the Ward--Takahashi identity.
Numerical results for different off--shell kinematics are presented.
\\
\\
PACS numbers: 11.30.R, 12.40, 13.40.F
\end{abstract}

\section{Introduction}
The electromagnetic form factors of strongly interacting particles play an
important role in investigations of electromagnetic processes involving
hadrons and virtual photons.
Furthermore, as they contain information about the underlying dynamics,
a good description of electromagnetic properties of hadrons serves
as a stringent test for phenomenological models of the strong interaction.
Assuming time--reversal invariance of the strong interaction, it is well--known
that the empirical description of free spin 0 particles requires one
single electromagnetic form factor, whereas a spin 1/2 system is described
by two electromagnetic form factors, e.g.\ a Dirac and a Pauli form factor.

However, it is very often not realized that, in general, a consistent treatment
of processes involving the electromagnetic interaction
will require information beyond the on--shell empirical form factors.
The structure of the electromagnetic vertex is more complicated
once the initial or final particle, or both, are not on the mass--shell.
In particular, more operator structures and thus more form factors appear.
Furthermore, the form factors will depend on additional scalar variables.

The off--shell electromagnetic structure of the nucleon
which requires, in general, twelve form factors \cite{Bincer}
has been investigated in several microscopic models over the past few years
\cite{Naus1,Tiemeijer,Song,Bos3}.
This interest was triggered by, among other things, the observation
that the interpretation of electron--nucleus scattering experiments
requires more than the information contained in the empirical on--shell
form factors, as the initial nucleon inside the nucleus is generally
not on its mass--shell.

The most general description of the pion electromagnetic vertex requires
two form factors $F$ and $G$ \cite{Nishijima,Barton} which are, however,
related by the Ward--Takahashi identity \cite{Ward,Takahashi}.
A systematic study of the off--shell vertex of the pion in a
microscopic model has not yet been performed.
This situation is somewhat surprising, as there has recently been quite
some discussion about investigating the ``pion content of the nucleon''
via pion electroproduction from the nucleon.
Clearly, in order to interpret the contribution of the pion in the t--channel
diagram of this process one has to address the question of how to describe
the interaction of particles not on their mass--shell.
Even close to threshold, the pion in the t--channel is off its mass--shell
by several units of its rest mass.
The possible importance of the inclusion of off--mass--shell
effects has been emphasized by Gross and Riska \cite{Gross}
in yet another context,
namely in the description of meson exchange currents, in which case
the exchanged pion is not on its mass--shell.
Finally, even in simple two--step processes, such as Compton scattering
from a {\em free} pion, a consistent description of the electromagnetic
interaction of particles which are not on their mass--shell is required,
as in the above process the intermediate pion in the pole diagrams
is not on its mass--shell.

Similar considerations apply to the kaon. At present, less is known about
the kaon but with the new generation of electron accelerators there
will be increased information about electromagnetic interactions of kaons.

In this work we will use chiral perturbation theory to $O(p^4)$
\cite{Weinberg,Gasser1,Gasser2} in order to discuss the most general
electromagnetic vertex of Goldstone bosons, which is consistent with
the requirements imposed by the chiral symmetry of the underlying
$QCD$ lagrangian.
Chiral perturbation theory has previously been used to determine
on--shell electromagnetic form factors \cite{Gasser3,Bijnens,Donoghue}.
In fact, the empirical value of the mean square radius of the pion
has been used as an input to fix one of the 10
phenomenological parameters of the $O(p^4)$ lagrangian, namely $L_9^r$
\cite{Gasser2,Gasser3,Bijnens}.
We will extend the analysis of Gasser and Leutwyler
\cite{Gasser1,Gasser2} to allow for the most general off--shell
Green's functions consistent with the requirements of chiral symmetry,
i.e., we identify the terms which originally have been eliminated by
making use of the classical equation of motion.
Two additional independent structures exist at order $p^4$,
one of which yields an off--shell contribution to the renormalized
electromagnetic vertex.
Even though chiral symmetry does not determine the coefficients of these
structures, it makes an unambiguous prediction in the sense
that the off--shell slope of the form factor $F$ of both, the charged
pion and kaon, is the same, namely it is proportional to one extra parameter.
Furthermore, the off--shell form factors of the $K^0$ are independent
of this parameter, and thus a unique prediction exists to order $p^4$
for this case.

Our paper is organized as follows. In the next section we will
introduce the concept of off--shell form factors.
In the third section we will briefly discuss the chiral lagrangian and
present the calculation of the form factors in chiral perturbation theory.
We will discuss our results and draw some conclusions in sect.\ 4.
In appendix A we will discuss in detail the Ward--Takahashi
identity and, finally, we will relate our convention with others used
in the literature.

\section{Definition of the off--shell form factors}

Starting point for the definition of the off--shell electromagnetic
form factors of, e.g., the $\pi^+$ is the three--point Green's function
\begin{equation}
\label{rtpgfc}
G^{\mu}(x,y,z)=<0|T\left(J^{\mu}(x) \pi^+(y) \pi^-(z)\right)|0>,
\end{equation}
where $J^{\mu}(x)$ is the electromagnetic current operator in units of
the elementary charge, $e>0$,
$\pi^{+/-}(x)$ destroys a $\pi^{+/-}$ or creates a $\pi^{-/+}$.
We define the Green's function in momentum space in terms of the Fourier
transform
\begin{equation}
\label{tpgfm}
(2\pi)^4 \delta^4(p_f-p_i-q) G^{\mu}(p_f,p_i)=
\int d^4x d^4y d^4z e^{-i(q \cdot x - p_f \cdot y  + p_i \cdot z )}
G^{\mu}(x,y,z),
\end{equation}
where $p_i, p_f$ are the four--momenta corresponding to the lines
entering and leaving the vertex, respectively, and $q=p_f-p_i$ is
the momentum transfer of the virtual photon at the vertex.
As usual, translational invariance enforces momentum conservation at the
vertex.
We introduce the renormalized three--point Green's function as \cite{Cheng}
\begin{equation}
\label{rtpgfm}
G^{\mu}_R(p_f,p_i) = Z^{-1}_{\phi} Z^{-1}_J G^{\mu}(p_f,p_i),
\end{equation}
where $Z_{\phi}$ and $Z_J$ are renormalization
constants\footnote{In fact, $Z_J=1$ due to gauge invariance \cite{Cheng}.}.
The irreducible, renormalized three--point Green's function is then
defined as
\begin{equation}
\label{irtpgf}
\Gamma^{\mu,irr}_R(p_f,p_i) =
(i \Delta_R(p_f))^{-1} G^{\mu}_R(p_f,p_i)(i\Delta_R(p_i))^{-1},
\end{equation}
where $\Delta_R(p)$ is the full, renormalized propagator.
In terms of Feynman diagrams $\Gamma^{\mu,irr}_R$ consists of contributions
which cannot be disconnected into two separate pieces by cutting one single
internal line.

For the most general parameterization of $\Gamma^{\mu,irr}_R$ we can form two
independent four--momenta, namely linear combinations of $p^{\mu}_i$ and
$p^{\mu}_f$, and three scalars, e.g., $q^2, p_i^2, p_f^2$.
Thus we may write \cite{Barton,Naus2} (the relation with other conventions
used is discussed in appendix B)
\begin{equation}
\label{par1}
\Gamma^{\mu,irr}_R(p_f,p_i) =
(p_f+p_i)^{\mu} F(q^2,p_f^2,p_i^2) +
(p_f-p_i)^{\mu} G(q^2,p_f^2,p_i^2).
\end{equation}
Assuming time--reversal invariance of the strong interaction we obtain
\begin{eqnarray}
\label{tri}
F(q^2,p_f^2,p_i^2)=F(q^2,p_i^2,p_f^2), \nonumber \\
G(q^2,p_f^2,p_i^2)=-G(q^2,p_i^2,p_f^2).
\end{eqnarray}
{}From eq.\ (\ref{tri}) we conclude that $G(q^2,p^2,p^2)=0$, in particular,
it vanishes for $p^2=M^2$, i.e., if the pion is on its mass shell.
Furthermore, using invariance of the strong interaction under charge
conjugation one finds that the electromagnetic form factors of
anti--particles are just the negative of those of particles.
Thus the $\pi^0$ and $\eta$ do not have any electromagnetic form factors
even off--shell.
In contrast, the $K^0$ (and similarly the $\bar{K}^0$) which is not its own
anti--particle may have form factors, just as the neutron does.

\section{Calculation of the off--shell form factors in chiral
perturbation theory}

In this section we will determine the most general off--shell
electromagnetic vertex in chiral perturbation theory to order $p^4$.
An extension of the standard formalism is required to describe the
most general off--shell Green's functions.
Naturally, the calculation is similar to the on--shell case
\cite{Gasser3,Bijnens,Donoghue} and thus some duplication of
already known results is unavoidable.

\subsection{The effective lagrangian}
In the original work by Gasser and Leutwyler use of the ``classical equation
of motion'', i.e., the one derived from the lowest--order effective lagrangian,
is made to eliminate structures at order $p^4$.
This procedure yields the most general result as long as one is concerned
with on--shell Green's functions.
As we are interested in the most general form of the Green's functions
even off--shell, we have to identify the additional structures which are
commonly abandoned.
For pedagogical reasons and for the sake of clarity, we find it necessary to
outline the construction of the most general lagrangian at order $p^4$.

We will only be concerned with the approximate chiral $SU(3)\times SU(3)$
symmetry of the $QCD$ lagrangian.
The 8 Goldstone bosons arising from the spontaneous symmetry breaking
are collected in a $SU(3)$ matrix
\begin{equation}
\label{u}
U(x)=\exp\left( i\frac{\phi(x)}{F_0} \right ),
\end{equation}
with
\begin{equation}
\label{Phi}
\phi(x)= \left (
\begin{array}{ccc}
\pi^0+\frac{1}{\sqrt{3}}\eta & \sqrt{2} \pi^+ & \sqrt{2} K^+ \\
\sqrt{2} \pi^- & -\pi^0+\frac{1}{\sqrt{3}}\eta & \sqrt{2} K^0 \\
\sqrt{2} K^- & \sqrt{2} \bar{K}^0 & -\frac{2}{\sqrt{3}}\eta
\end{array}
\right ),
\end{equation}
and $F_0$ the pseudoscalar meson decay constant in the chiral limit
\cite{Gasser2}.
The matrix $U$ transforms linearly under the group $G=SU(3)_L \times SU(3)_R:
U \rightarrow U'=V_R U V_L^{\dagger}$, where $V_L$ and $V_R$ are independent
$SU(3)$ matrices.
In order to promote the global symmetry to a local symmetry with respect to $G$
one introduces the covariant derivative of $U$,
and in fact of any operator $A$ which transforms as $V_R A V_L^{\dagger}$.
For that purpose one needs 16 external gauge fields
$L^a_{\mu}$ and $R^a_{\mu}$ which are collected in traceless hermitian
$3\times 3$ matrices, $L_{\mu}=\frac{\lambda^a}{2}L^a_{\mu}$ and
$R_{\mu}=\frac{\lambda^a}{2}R^a_{\mu}$.
Finally, one introduces field strength tensors $F^L_{\mu\nu}$ and
$F^R_{\mu\nu}$ for the gauge fields and
external scalar and pseudoscalar sources $s$ and $p$.
The corresponding transformation properties read \cite{Gasser2}
\begin{eqnarray}
\label{traf}
D_{\mu} U  \equiv  \partial_{\mu}U-iR_{\mu}U+iU L_{\mu}
& \stackrel{G}{\rightarrow} &
V_R D_{\mu} U V_L^{\dagger}, \nonumber \\
R_{\mu} & \stackrel{G}{\rightarrow} & V_R R_{\mu} V_R^{\dagger}
+i V_R \partial_{\mu} V_R^{\dagger}, \nonumber \\
L_{\mu} & \stackrel{G}{\rightarrow} & V_L L_{\mu} V_L^{\dagger}
+i V_L \partial_{\mu} V_L^{\dagger}, \nonumber \\
F_{\mu\nu}^R \equiv\partial_{\mu}R_{\nu}-\partial_{\nu}R_{\mu}-
i[R_{\mu},R_{\nu}]
& \stackrel{G}{\rightarrow} & V_R F_{\mu\nu}^R V_R^{\dagger}, \nonumber \\
F_{\mu\nu}^L \equiv\partial_{\mu}L_{\nu}-\partial_{\nu}L_{\mu}-
i[L_{\mu},L_{\nu}]
& \stackrel{G}{\rightarrow} & V_L F_{\mu\nu}^L V_L^{\dagger}, \nonumber \\
\chi\equiv 2B_0(s+ip) & \stackrel{G}{\rightarrow} & V_R \chi V_L^{\dagger},
\end{eqnarray}
where $B_0$ is a constant introduced for convenience which is related
to the vacuum expectation value $<0|\bar{q}q|0>$
(see e.g.\ ref.\ \cite{Donoghue} for further details).

In the power counting scheme of chiral perturbation theory the above terms
are booked as:
\begin{equation}
\label{powercounting}
U =  O(p^0),\, D_{\mu} U  =  O(p),\, R_{\mu},L_{\mu}  =  O(p),\,
F^{L/R}_{\mu\nu}  =  O(p^2),\, \chi  =  O(p^2).
\end{equation}
The construction of the effective lagrangian in terms of the building blocks
of eq.\ (\ref{traf}) proceeds as follows.
Given operators $A,B,\dots$, all of which transform as
\mbox{$A'=V_R A V_L^{\dagger},$} \mbox{$B'= V_R B V_L^{\dagger},\,\dots,$}
one can form invariants by taking the trace of products of the
type $A B^{\dagger}$.
The generalization to more terms is obvious and, of course, the product of
invariant traces is invariant.

Any operator $A$ of order $p^4$, transforming as $A'=V_R A V_L^{\dagger}$,
can be expressed in the form $A_1, A_1 A_2^{\dagger} A_3, \dots,$
where the $A_i$ are elements of the following list:
\begin{eqnarray}
\label{list1}
& & U, D_{\mu} U, D_{\mu} D_{\nu}U, D_{\mu} D_{\nu} D_{\rho} U,
D_{\mu} D_{\nu} D_{\rho} D_{\sigma} U, \nonumber \\
& & \chi, D_{\mu} \chi, D_{\mu} D_{\nu} \chi, \nonumber \\
& & U F^L_{\mu\nu},D_{\rho}(U F^L_{\mu\nu}),
D_{\rho} D_{\sigma} (U F^L_{\mu\nu}),\nonumber \\
& & F^R_{\mu\nu} U ,D_{\rho}(F^R_{\mu\nu} U),
D_{\rho} D_{\sigma} (F^R_{\mu\nu} U).
\end{eqnarray}
The key to this statement is the ``chain rule'',
\begin{eqnarray}
\label{cr}
D_{\mu}(A_1 A_2^{\dagger}A_3...A_{2n+1})&=&
D_{\mu}A_1 A_2^{\dagger}A_3...A_{2n+1}+
A_1 (D_{\mu}A_2)^{\dagger} A_3...A_{2n+1} \nonumber\\
&&+... +A_1 A_2^{\dagger}A_3...D_{\mu}A_{2n+1},\quad (n\geq 1),
\end{eqnarray}
which is easily verified via induction using the definition of the covariant
derivative.
As an example, one has
\begin{eqnarray}
\label{excr}
D_{\mu}\left(U(D_{\nu}U)^{\dagger}D_{\rho}U\right) &=&
D_{\mu}U (D_{\nu}U)^{\dagger} D_{\rho}U+ \nonumber \\&&
U (D_{\mu}D_{\nu}U)^{\dagger} D_{\rho}U+
U (D_{\nu}U)^{\dagger} D_{\mu}D_{\rho}U.
\end{eqnarray}

In fact, not all operators of eq.\ (\ref{list1}) are needed
due to the following ``total derivative argument''.
For any pair of operators $A$ and $B$ transforming as $V_R...V_L^{\dagger}$
one finds
\begin{equation}
\label{pr}
D_{\mu}A B^{\dagger}+A (D_{\mu}B)^{\dagger} = \partial_{\mu}(A B^{\dagger})
-i[R_{\mu},A B^{\dagger}],
\end{equation}
and thus using the fact that the trace of a commutator vanishes,
\begin{equation}
\label{rm}
Tr\left(D_{\mu}A B^{\dagger}\right)=
-Tr\left(A (D_{\mu}B)^{\dagger}\right)+
\partial_{\mu}Tr\left(A B^{\dagger}\right).
\end{equation}
With the help of eq.\ (\ref{rm}) covariant derivatives can be moved in
single--trace expressions by introducing total derivatives which do not
contribute to the equation of motion and thus may be dropped.
This technique may be used at order $p^2$ to remove terms with two
covariant derivatives\footnote{At order $p^2$ we could also use
$Tr(D_{\mu}D_{\nu}U U^{\dagger})=-Tr(D_{\nu}U(D_{\mu}U)^{\dagger})$.}
on $U$, and at order $p^4$ to eliminate
covariant derivatives of expressions which are already order $p^2$.
Thus our list of building blocks effectively reduces to
\begin{eqnarray}
\label{list2}
&&U, D_{\mu} U, \chi, \quad \mbox{at $O(p^2)$,}\nonumber \\
&&U, D_{\mu} U, D_{\mu} D_{\nu}U, \chi, U F^L_{\mu\nu}, F^R_{\mu\nu} U,
\quad \mbox{at $O(p^4)$.}
\end{eqnarray}
In eq. (\ref{list2}) we omitted the field strength tensors at order $p^2$
as they vanish upon contraction of the Lorentz indices.
The number of invariants is further reduced by observing
\begin{equation}
U (D_{\mu} U)^{\dagger}=-D_{\mu}U U^{\dagger}, \quad
Tr\left(U (D_{\mu} U)^{\dagger}\right)=0,
\end{equation}
which restricts the invariant structures through order $p^4$ to the type
$Tr(O(p^2))$, $Tr(O(p^4))$, and $Tr(O(p^2))Tr(O(p^2))$, where we omit
an irrelevant constant at order $p^0$.
Forming Lorentz--invariant and parity--even combinations
the most general lagrangian at order $p^2$ is given by \cite{Gasser2}
\begin{equation}
\label{l2}
{\cal L}_2 = \frac{F_0^2}{4} Tr \left ( D_{\mu} U (D^{\mu}U)^{\dagger} \right)
+\frac{F_0^2}{4} Tr \left ( \chi U^{\dagger}+ U \chi^{\dagger} \right ).
\end{equation}
The equation of motion derived from the lowest--order lagrangian
${\cal L}_2$ is
\begin{equation}
\label{eom}
{\cal O}=D_{\mu}D^{\mu}U U^{\dagger}-U(D_{\mu}D^{\mu}U)^{\dagger}
-\chi U^{\dagger}+U\chi^{\dagger}
+\frac{1}{3}Tr\left(\chi U^{\dagger}-U\chi^{\dagger}\right)=0.
\end{equation}
For later use we have introduced $\cal O$ as an abbreviation for the
expression in the lowest--order equation of motion.
The origin of the trace term is the constraint on the $U$ matrix, $det(U)=1$.
It ensures that eq.\ (\ref{eom}) represents 8 independent equations of motion
(8 Goldstone bosons!) and not 9 as one might naively expect from a
$3\times 3$ matrix equation.

Using in addition charge conjugation and a relation among $SU(3)$ generators
\cite{comment1} the $O(p^4)$ lagrangian can be cast into the form  given by
Gasser and Leutwyler in ref.\ \cite{Gasser2}
\begin{eqnarray}
\label{l4gl}
\lefteqn{{\cal L}^{G\&L}_4  =
L_1 \left ( Tr(D_{\mu}U (D^{\mu}U)^{\dagger}) \right)^2
+ L_2 Tr \left (D_{\mu}U (D_{\nu}U)^{\dagger}\right)
Tr \left (D^{\mu}U (D^{\nu}U)^{\dagger}\right)}
\nonumber \\
& & + L_3 Tr \left (D_{\mu}U (D^{\mu}U)^{\dagger}D_{\nu}U (D^{\nu}U)^{\dagger}
\right )\nonumber\\
& & + L_4 Tr \left ( D_{\mu}U (D^{\mu}U)^{\dagger} \right )
Tr \left( \chi U^{\dagger}+ U \chi^{\dagger} \right )
\nonumber \\
& & +L_5 Tr \left( D_{\mu}U (D^{\mu}U)^{\dagger}
(\chi U^{\dagger}+ U \chi^{\dagger})\right)
+ L_6 \left( Tr \left ( \chi U^{\dagger}+ U \chi^{\dagger} \right )
\right)^2
\nonumber \\
& & + L_7 \left( Tr \left ( \chi U^{\dagger} - U \chi^{\dagger} \right )
\right)^2
+ L_8 Tr \left ( U \chi^{\dagger} U \chi^{\dagger}
+ \chi U^{\dagger} \chi U^{\dagger} \right )
\nonumber \\
& & -i L_9 Tr \left ( F^R_{\mu\nu} D^{\mu} U (D^{\nu} U)^{\dagger}
+ F^L_{\mu\nu} (D^{\mu} U)^{\dagger} D^{\nu} U \right )
+ L_{10} Tr \left ( U F^L_{\mu\nu} U^{\dagger} F_R^{\mu\nu} \right )
\nonumber \\
& & + H_1 Tr \left ( F^R_{\mu\nu} F^{\mu\nu}_R +
F^L_{\mu\nu} F^{\mu\nu}_L \right )
+ H_2 Tr \left ( \chi \chi^{\dagger} \right ).
\end{eqnarray}
The parameters $L_i$ are not determined by chiral symmetry.
They can either be determined empirically by fitting experimental
results \cite{Gasser2}, or predictions may be derived from $QCD$ inspired
effective quark models.
Some of the $L_i$ are infinite, i.e., they are used to renormalize
1--loop diagrams from ${\cal L}_2$ (see refs.\ \cite{Gasser1,Gasser2,Donoghue}
for further details).
In the derivation of eq.\ (\ref{l4gl}) the equation of motion of
${\cal L}_2$, eq.\ (\ref{eom}), was used to eliminate two additional
terms \cite{comment2}.
Using the expressions of eq.\ (\ref{list2}) one finds that these
structures contain two covariant derivatives acting on $U$ and may,
for example, be written as
\begin{equation}
\label{addstr}
Tr\left(D_{\mu}D^{\mu}U (D_{\nu}D^{\nu}U)^{\dagger}\right),\,
Tr\left(D_{\mu}D^{\mu}U \chi^{\dagger}+\chi (D_{\mu}D^{\mu}U)^{\dagger}\right).
\end{equation}
In order to keep the same values of the original coefficients $L_i$
on-- and off--shell we choose instead of the structures
of eq.\ (\ref{addstr}) the following combinations
\begin{equation}
\label{los}
{\cal L}^{off-shell}_4=
\beta_1 Tr\left({\cal O}{\cal O}^{\dagger}\right)
+\beta_2Tr\left((\chi U^{\dagger}-U\chi^{\dagger}){\cal O}\right),
\end{equation}
which vanish identically when the on--shell equation of motion
${\cal O}=0$ is used.
It is tedious but straightforward to show that eq.\ (\ref{los}) can
be expressed in terms of the structures of eqs.\ (\ref{l4gl}) and
(\ref{addstr}).

\subsection{The electromagnetic current operator}
In order to obtain the electromagnetic current operator corresponding to
the effective lagrangians of eqs.\ (\ref{l2}), (\ref{l4gl}) and (\ref{los})
we first identify the external fields with
\begin{eqnarray}
\label{lra}
L^{\mu}=R^{\mu}=-e Q A^{\mu},\quad && Q=
\left(
\begin{array}{ccc}
2/3 & 0 & 0 \\
0 & -1/3 & 0 \\
0 & 0 & -1/3
\end{array}
\right ),\nonumber \\
\chi = 2 B_0 M, \quad && M=
\left(
\begin{array}{ccc}
m_u & 0 & 0 \\
0 & m_d & 0 \\
0 & 0 & m_s
\end{array}
\right ),
\end{eqnarray}
where $Q$ is the quark charge matrix and $M$ is the quark mass matrix.
In the following we will restrict ourselves to the isospin symmetric
limit $m_u=m_d=m$.
With these definitions, the covariant derivative and the
field strength tensors of eqs.\ (\ref{traf}) become
\begin{equation}
D_{\mu}U = \partial_{\mu} U + i e A_{\mu}[Q,U], \quad
F_{\mu\nu}^R=F_{\mu\nu}^L=-e(\partial_{\mu}A_{\nu}-\partial_{\nu}A_{\mu})Q.
\end{equation}
The current operator then results from taking the derivative
\begin{equation}
J^{\mu}=\left. -\frac{1}{e}\frac{\partial {\cal L}}{\partial A_{\mu}}
\right|_{A=0}.
\end{equation}

The renormalized 3--point Green's function, $\Gamma^{\mu,irr}_R(p_f,p_i)$,
is calculated perturbatively using standard
techniques (see e.g.\ ref.\ \cite{Bjorken}, chpt.\ 17.3).
Applying the power counting scheme of chiral perturbation theory
\cite{Weinberg} (see e.g.\ ref.\ \cite{Donoghue} p.\ 108) one can see that the
${\cal L}_2$
lagrangian will contribute with $p^2$ at tree--diagram level, and with $p^4$
at one--loop level\footnote{We book the polarization vector of the photon
as $O(p)$.}
(see figs.\ 1,2,3).
The ${\cal L}_4$ lagrangian will contribute at $O(p^4)$ at tree level
(see fig.\ 4)
and at least at $O(p^6)$ in loop--diagrams, which thus are to be
neglected here.

\subsection{Tree--level contribution from ${\cal L}_2$ and ${\cal L}_4$}

The tree--level diagrams derived from ${\cal L}_2$ obtain contributions
from the piece of the current operator which is of second order in the
Goldstone boson fields,
\begin{equation}
\label{j22}
J^{\mu,2}_2=-\frac{i}{2}Tr\left(Q[\phi,\partial^{\mu}\phi]\right)
=i(\pi^-\partial^{\mu}\pi^+-\pi^+\partial^{\mu}\pi^-)
+i(K^-\partial^{\mu}K^+-K^+\partial^{\mu}K^-).
\end{equation}
The superscript 2 denotes the expansion to second order in $\phi$
and the subscript indicates that the operator is derived from
${\cal L}_2$.
In terms of diagrams, eq.\ (\ref{j22}) gives rise to the usual
pointlike convection current, $(p_f+p_i)^{\mu}$, for positively
charged pions and kaons (see fig.\ 1).

Expanding the ${\cal L}_4$ part of the Gasser and Leutwyler lagrangian
(see eq.\ (\ref{l4gl})) we find that only the terms with
coefficients $L_4, L_5$ and $L_9$ have an $O(\phi^2)$ component
also containing $e A_{\mu}$.
After some algebra one obtains for the $O(p^4)$ current operator:
\begin{eqnarray}
\label{j42}
J^{\mu,2}_4&=&i c_{\pi}(\pi^-\partial^{\mu}\pi^+-\pi^+\partial^{\mu}\pi^-)
+i c_K (K^-\partial^{\mu}K^+-K^+\partial^{\mu}K^-) \nonumber \\
& & -4i \frac{L_9}{F^2} \partial_{\nu}
(\partial^{\nu}\pi^-\partial^{\mu}\pi^+
-\partial^{\nu}\pi^+\partial^{\mu}\pi^-) \nonumber \\
& & -4i \frac{L_9}{F^2} \partial_{\nu}
(\partial^{\nu}K^-\partial^{\mu}K^+
-\partial^{\nu}K^+\partial^{\mu}K^-),
\end{eqnarray}
where the (infinite) constants $c_{\pi}$ and $c_K$ are given by
\begin{eqnarray}
\label{cs}
c_{\pi}&=&8 \frac{(2 M^2_{\pi}+3M^2_{\eta}-2M^2_K)L_4+M^2_{\pi}L_5}{F^2},
\nonumber \\
c_{K}&=&8 \frac{(2 M^2_{\pi}+3M^2_{\eta}-2M^2_K)L_4+M^2_{K}L_5}{F^2}.
\end{eqnarray}
In order to arrive at eq.\ (\ref{cs}), we made replacements of the type
\mbox{$(m_u+m_d)B_0$} \mbox{$\rightarrow M^2_{\pi}$},
\mbox{$F_0\rightarrow F \approx 93 MeV$}.
Such replacements are allowed in expressions which are already
$O(p^4)$ in leading order, such as e.g.\ eq.\ (\ref{j42}),
without changing the results at $O(p^4)$.
Furthermore, when deriving the term proportional to $L_9$ in eq.\ (\ref{j42}),
we made use of
\begin{equation}
\label{totder}
\partial_{\nu}A_{\mu} f=\partial_{\nu}(A_{\mu}f)-A_{\mu}\partial_{\nu}f,
\end{equation}
where $f$ is an arbitrary function of the fields and their derivatives.
We then dropped the total derivative on the right--hand side of
eq.\ (\ref{totder}), as it does not change the equation of motion.

Finally, the two additional structures which are not contained in the original
work of Gasser and Leutwyler lead to the following operators:
\begin{eqnarray}
\label{j42off}
J^{\mu,2}_{4,off,a}&=&
-\frac{16i \beta_1}{F^2}\Big(
(\Box+M^2_{\pi})\pi^-\partial^{\mu}\pi^+
+\pi^-(\Box+M^2_{\pi})\partial^{\mu}\pi^+
\nonumber \\ &&
-(\Box+M^2_{\pi})\pi^+\partial^{\mu}\pi^-
-\pi^+(\Box+M^2_{\pi})\partial^{\mu}\pi^-
\nonumber \\ &&
+(\Box+M^2_K)K^-\partial^{\mu}K^+
+K^-(\Box+M^2_K)\partial^{\mu}K^+
\nonumber \\ &&
-(\Box+M^2_K)K^+\partial^{\mu}K^-
-K^+(\Box+M^2_K)\partial^{\mu}K^-\Big),
\nonumber \\
J^{\mu,2}_{4,off,b}&=&
-16i \beta_2 \frac{M^2_{\pi}}{F^2}
(\pi^-\partial^{\mu}\pi^+-\pi^+\partial^{\mu}\pi^-) \nonumber \\
& & -16i \beta_2 \frac{M^2_K}{F^2}
(K^-\partial^{\mu}K^+-K^+\partial^{\mu}K^-).
\end{eqnarray}
In deriving eq.\ (\ref{j42off}), we once again made use of
the total derivative argument to remove derivatives acting on the
four--potential $A_{\mu}$.

It is interesting to note that all tree--level terms are of the
``minimal substitution type'',
i.e., there are direct couplings to charged particles only.
This is nontrivial in the sense, that even though the $K_0$ has
no charge, it has a charge distribution, and thus an effective
coupling involving higher order derivatives in principle could
have been possible.

Applying standard Feynman rules, the current operators
of eqs.\ (\ref{j22}), (\ref{j42}) and (\ref{j42off}) result
in the following contribution to the yet unrenormalized vertex
(see figs.\ 1 and 4):
\begin{eqnarray}
\label{fd4}
\Gamma^{\mu}_{tree}(\pi^+)&=&(p_f+p_i)^{\mu}
\left(1+d_{\pi}+2L_9\frac{q^2}{F^2}
+16\beta_1 \frac{p_f^2+p^2_i-2M^2_{\pi}}{F^2}\right)\nonumber\\
&&+(p_f-p_i)^{\mu}2L_9\frac{p_i^2-p_f^2}{F^2}, \nonumber \\
\Gamma^{\mu}_{tree}(K^+)&=&(p_f+p_i)^{\mu}
\left(1+d_K+2L_9\frac{q^2}{F^2}
+16\beta_1\frac{p_f^2+p^2_i-2M^2_K}{F^2}\right)\nonumber\\
&&+(p_f-p_i)^{\mu}2L_9\frac{p_i^2-p_f^2}{F^2},
\end{eqnarray}
where
\begin{equation}
\label{ds}
d_{\pi}=c_{\pi}-16\beta_2\frac{M^2_{\pi}}{F^2}, \quad
d_K=c_K-16\beta_2\frac{M^2_K}{F^2}.
\end{equation}
It is worthwhile to note that we get two kinds of off--shell contributions.
The $L_9$ term enters into the $G$ form factor, whereas the $\beta_1$
term contributes to the off--shell behavior of the unrenormalized $F$ form
factor.
The $\beta_2$ term appears only in $d_{\pi}$ and $d_K$ which will be removed
by wave function renormalization.

\subsection{One--loop contributions}

We now turn to the one--loop diagrams involving interaction vertices derived
from ${\cal L}_2$. These diagrams can be grouped into two different classes.
The first involves an interaction vertex with one photon and four
Goldstone bosons, two of which will be contracted to form a
loop (1--vertex loop diagram, see fig.\ 2).
The second consists of a contraction of the current operator of
eq.\ (\ref{j22}) with an interaction vertex containing four Goldstone bosons
(see fig.\ 3).
Clearly, in the second case, only charged particles can contribute
in the loop.

\subsubsection{1--vertex loop diagram}

The expression for the current operator with four Goldstone boson fields
is too complicated to be given explicitly in terms of the charged states
and thus we will quote the result in terms of $\phi$ only,
\begin{equation}
\label{j24}
J^{\mu,4}_2=\frac{i}{24F_0^2}Tr\Big(Q(\phi^2[\phi,\partial^{\mu}\phi]
-2\phi [\phi,\partial^{\mu}\phi] \phi
+[\phi,\partial^{\mu}\phi]\phi^2)\Big).
\end{equation}
Contracting two fields in eq.\ (\ref{j24}) to form a loop, will
result in infinities which are isolated using the method of dimensional
regularization.
Note that contractions of a field with its derivative at the same
space--time point vanish.
The contribution to the unrenormalized vertex of the charged pion reads,
\begin{equation}
\label{tadpolecontrpi}
\Gamma^{\mu}_{tad}(\pi^+)=(p_f+p_i)^{\mu}
\left( -\frac{5}{3} \frac{I(M_{\pi}^2,\mu^2)}{F^2}
-\frac{5}{6} \frac{I(M_K^2,\mu^2)}{F^2}\right ),
\end{equation}
where $I(M^2,\mu^2)$ results from the contraction of two fields at the same
space--time point and is defined as
\begin{eqnarray}
\label{im}
I(M^2,\mu^2)&=&\mu^{4-d}\int \frac{d^d k}{(2\pi)^d} \frac{i}{k^2-M^2+i\epsilon}
\nonumber \\
&=& \frac{M^2}{16\pi^2}\left(R+
\ln\left[\frac{M^2}{\mu^2}\right]\right) + O(4-d),
\end{eqnarray}
with $\mu$ the renormalization scale.
We use the same convention as in ref.\ \cite{Donoghue}, p.\ 169 for the
dimensional regularization.
In eq.\ (\ref{im}) we introduced the abbreviation
\begin{equation}
\label{r}
R=\frac{2}{d-4}-(\ln(4\pi)+\Gamma'(1)+1).
\end{equation}
The first term in eq.\ (\ref{tadpolecontrpi}) results from a pion
in the loop and the second one from a kaon.

The corresponding contribution for the kaon is given by
\begin{eqnarray}
\label{tadpolecontrkaon}
\Gamma^{\mu}_{tad}(K^+) &=&(p_f+p_i)^{\mu}
\left( -\frac{1}{4} \frac{I(M_{\eta}^2,\mu^2)}{F^2}
-\frac{3}{4} \frac{I(M_{\pi}^2,\mu^2)}{F^2}
-\frac{3}{2} \frac{I(M_K^2,\mu^2)}{F^2}\right ), \nonumber\\
\Gamma^{\mu}_{tad}(K^0)&=&(p_f+p_i)^{\mu}
\left( \frac{1}{2} \frac{I(M_{\pi}^2,\mu^2)}{F^2}
-\frac{1}{2} \frac{I(M_K^2,\mu^2)}{F^2}\right ).
\end{eqnarray}
Notice that in the case of the charged kaon the eta appears in the loop
as well.

\subsubsection{2--vertex loop diagram}

A second class of loop diagrams results from contracting the current operator
of eq.\ (\ref{j22}), $J^{\mu,2}_2$, with  the piece in ${\cal L}_2$ which
describes the scattering of 2 Goldstone bosons and which is given by
\begin{equation}
\label{lscatt}
{\cal L }^4_2=\frac{1}{24F^2_0}\left(Tr([\phi,\partial_{\mu}\phi]
\phi \partial^{\mu}\phi)+B_0Tr(\phi^4 M)\right).
\end{equation}
Notice that the internal lines of the resulting Feynman diagrams
(see fig.\ 3) must be charged lines.
Thus we only require the following Feynman rules
derived from eq.\ (\ref{lscatt}):
\begin{eqnarray}
\label{iv}
(\pi^+,\pi^-|\pi^+,\pi^-)&:&
\frac{i}{3F^2}\left(2M^2_{\pi}+g(p_1,p_2,k_1,k_2)\right),
\nonumber\\
(\pi^+,K^-|\pi^+,K^-)&:&
\frac{i}{6F^2}\left(M^2_{\pi}+M^2_K+g(p_1,p_2,k_1,k_2)\right),
\nonumber\\
(K^+,\pi^-|K^+,\pi^-)&:&
\frac{i}{6F^2}\left(M^2_{\pi}+M^2_K+g(p_1,p_2,k_1,k_2)\right),
\nonumber\\
(K^+,K^-|K^+,K^-)&:&
\frac{i}{3F^2}\left(2M^2_K+g(p_1,p_2,k_1,k_2)\right),
\nonumber\\
(K^0,\pi^-|K^0,\pi^-)&:&
\frac{i}{6F^2}\left(M^2_{\pi}+M^2_K+g(p_1,-k_2,k_1,-p_2)\right),
\nonumber\\
(K^0,K^-|K^0,K^-)&:&
\frac{i}{6F^2}\left(2M^2_K+g(p_1,p_2,k_1,k_2)\right),
\end{eqnarray}
where $(A,B|A,B)$ stands for $A(p_1)+B(p_2)\rightarrow A(k_1)+B(k_2)$
and the function $g$ is defined as
\begin{equation}
\label{gfunction}
g(p_1,p_2,k_1,k_2)=2p_1\cdot k_2+2k_1\cdot p_2+(p_1-k_2)\cdot(p_2-k_1).
\end{equation}
Note the overall factor of two in the case of $\pi^+\pi^-$ and
$K^+ K^-$ scattering as well as the arguments of the function $g$ for
$K^0 \pi^-$ scattering\footnote{In eq.\ (\ref{iv}) we have written $F$ and $M$
instead of the lowest--order predictions, since we will
use it as a building block of a higher--order diagram.}.

For example, the pion--loop diagram of fig.\ 5 contributes in
the following way to the vertex operator of the pion:
\begin{eqnarray}
\label{loopint}
\int \frac{d^4k}{(2\pi)^4} i\Delta_{M_{\pi}^2}(q+k) i\Delta_{M_{\pi}^2}(k)
(-1)(2k+q)^{\mu} \frac{i}{3F^2}\left(2M^2_{\pi}+g(p_i,q+k,p_f,k)\right)
&& \nonumber \\
\rightarrow \frac{2i\mu^{4-d}}{F^2}\int \frac{d^d k}{(2\pi)^d}
\frac{k^{\mu}k\cdot(p_f+p_i)}{[(k+\frac{1}{2}q)^2-M^2_{\pi}]
[(k-\frac{1}{2}q)^2-M^2_{\pi}]}, \nonumber \\
\end{eqnarray}
where $\Delta_{M^2}(p)$ is the free propagator of a particle with mass
$M$.
In eq.\ (\ref{loopint}) we assumed that the integral converges with a
suitable choice of dimension $d$ in order to be able to shift variables
and make use of a symmetric--integration argument.
The calculation of the integral of eq.\ (\ref{loopint}) is outlined
in ref.\ \cite{Donoghue}.
Note, however, that we have to keep the contribution proportional to $q^{\mu}$
which was omitted in the on--shell calculation of ref.\ \cite{Donoghue}.
Using standard techniques (dimensional regularization and Feynman
parameterization) the results of the 2--vertex loop diagrams are given by
\begin{eqnarray}
\label{looppi}
\Gamma^{\mu}_{loop}(\pi^+)&=& \Bigg \{ (p_f+p_i)^{\mu}
\left( \frac{I(M^2_{\pi},\mu^2)}{F^2}+q^2 \left( 2 A(q^2,M^2_{\pi},\mu^2)-
\frac{R}{96\pi^2F^2} \right) \right)
\nonumber \\
&&+(p_f-p_i)^{\mu}(p_i^2-p_f^2)\left(2 A(q^2,M^2_{\pi},\mu^2)
-\frac{R}{96\pi^2 F^2}\right)\Bigg\}\nonumber\\
&&+\frac{1}{2}\Bigg\{ M^2_{\pi}\rightarrow M^2_K \Bigg \}, \\
\label{loopk0}
\Gamma^{\mu}_{loop}(K^0)&=& \Bigg \{ (p_f+p_i)^{\mu}
\left(-\frac{I(M^2_{\pi},\mu^2)}{2F^2}-q^2 \left(  A(q^2,M^2_{\pi},\mu^2)-
\frac{R}{192\pi^2F^2} \right) \right)
\nonumber \\
&&+(p_f-p_i)^{\mu}(p_i^2-p_f^2)\left(-A(q^2,M^2_{\pi},\mu^2)
+\frac{R}{192\pi^2 F^2}\right)\Bigg\}\nonumber\\
&&-\Bigg\{ M^2_{\pi}\rightarrow M^2_K \Bigg \},
\end{eqnarray}
where it is convenient to introduce the function
\begin{equation}
\label{afunction}
A(q^2,M^2,\mu^2)=-\frac{1}{192 \pi^2 F^2}\left(\ln\left(\frac{M^2}{\mu^2}
\right)
+\frac{1}{3}+\left(1-4 \frac{M^2}{q^2}\right)H\left(\frac{q^2}{M^2}\right)
\right),
\end{equation}
and the function $H(a)$ is defined as\footnote{Note the difference in
comparison with eq.\ (3.10) of ref.\ \cite{Donoghue} which seems to
contain typographical errors.}
\begin{eqnarray}
\label{ha}
H(a)&=&\int_0^1 dx \ln \left(1+a(x^2-x)\right)\nonumber\\
&=& \left \{ \begin{array}{l}-2+2\sqrt{\frac{4}{a}-1}\,\mbox{arccot}
\left(\sqrt{\frac{4}{a}-1}\right)\quad (0<a<4),\\
-2+\sqrt{1-\frac{4}{a}}\left(\ln\left(\left|
\frac{\sqrt{1-\frac{4}{a}}+1}{\sqrt{1-\frac{4}{a}}-1}\right|\right)
+i\pi\Theta(a-4)\right) \quad \mbox{(otherwise).}\end{array} \right .
\nonumber \\
\end{eqnarray}
In the next section we will see that the form factors are written in
terms of only the function $A$ and the phenomenological constants.
The expression for $\Gamma^{\mu}_{loop}(K^+)$ can be obtained from
$\Gamma^{\mu}_{loop}(\pi^+)$ by interchanging $M^2_{\pi}$ and
$M^2_K$.
This simple rule is a result of the similarity of the vertices of
eq.\ (\ref{iv}).

\subsection{Renormalization}

The result for the renormalized vertex of the $\pi^+$ is obtained by
adding up the contributions of eqs.\ (\ref{fd4}),
(\ref{tadpolecontrpi}), (\ref{looppi}) and multiplying the result
by the
wave function renormalization constant $Z_{\pi}$ (and similarly for the $K^+$).
In fact, as $Z_{\pi}$ is of the form $Z_{\pi}=1+O(p^2)$
(see eq.\ (\ref{wfrcs}) below),
it is only the tree--level contribution derived from ${\cal L}_2$ which gets
modified \cite{Donoghue}.
The situation for the $K^0$ is simpler as there is no tree--level
contribution.
Thus, to $O(p^4)$ the result is already given by the sum of
eqs.\ (\ref{tadpolecontrkaon}) and (\ref{loopk0}).

The wave function renormalization constants are obtained in the standard
fashion \cite{Cheng} in terms of the unrenormalized self energy
$\Sigma_{\phi} (p^2)$,
\begin{equation}
\label{wfrc}
Z_{\phi}=\frac{1}{1-\Sigma'_{\phi}(M^2_{\phi})}.
\end{equation}
For the standard Gasser and Leutwyler approach, the self energy is of the form
\begin{equation}
\label{se}
\Sigma_{\phi}(p^2)=A_{\phi}+B_{\phi}p^2,
\end{equation}
which can be understood in the following way.
In terms of diagrams, to $O(p^4)$ the self energy results from the
one--loop diagram of fig.\ 6 and the tree--level diagram of
fig.\ 7.
The ${\cal L}_2$ lagrangian contains either no or two derivatives of the
fields and the interaction piece which contributes to the one--loop diagram
is symbolically either of the type $\phi^4$ or
$\phi\phi\partial\phi\partial\phi$.
The first type does not give rise to any $p^2$ dependence (no derivatives!)
whereas the second type results in terms proportional to either $M^2_{\phi}$
if the $\phi$'s are connected to the external legs, or $p^2$ if the
$\partial\phi$'s are connected to the external legs.
Even though the ${\cal L}_4$ lagrangian may contain four derivatives
($L_1,L_2,L_3$ terms) such terms do not contribute at tree level as
they are of fourth order in the fields. Thus also here we have either
a $p^2$ ($L_4,L_5$) or a constant ($L_6,L_8$ and $L_7$ for the $\eta$ only)
contribution.

The contribution of the additional terms of eqs.\ (\ref{los}) to the self
energy is
\begin{equation}
\label{seos}
\Sigma^{off}_{\phi}(p^2)=-16\beta_1 \frac{(p^2-M^2_{\phi})^2}{F^2}
+16\beta_2 \frac{M^2_{\phi}(p^2-M^2_{\phi})}{F^2},
\end{equation}
where we replaced all relevant quantities by their $O(p^4)$ prediction
as the difference is of yet higher order.
With the help of eq.\ (\ref{wfrc}) one finds for the relevant wave function
renormalization constants
\begin{eqnarray}
\label{wfrcs}
Z_{\pi}&=&1-d_{\pi}+\frac{2I(M^2_{\pi},\mu^2)}{3F^2}
+\frac{I(M^2_K,\mu^2)}{3F^2}+O(p^4), \nonumber\\
Z_{K}&=&1-d_K+\frac{I(M^2_{\pi},\mu^2)}{4F^2}
+\frac{I(M^2_K,\mu^2)}{2F^2}+\frac{I(M^2_{\eta},\mu^2)}{4F^2}+O(p^4),
\end{eqnarray}
where $d_{\pi}$ and $d_K$ are defined by eqs.\ (\ref{cs}) and (\ref{ds}).

Finally, we have to determine the renormalized propagator in order to
verify that our results satisfy the Ward--Takahashi identity.
Owing to the simple form of the self energy of eq.\ (\ref{se})
the renormalized propagator in the standard Gasser and Leutwyler
approach is given by
\begin{eqnarray}
\label{rpgl}
i\Delta_R^{G\&L}(p)&=&
\frac{i Z_{\phi}^{-1}}{p^2-M_0^2-\Sigma_{\phi}(p^2)+i\epsilon}\nonumber \\
&=&\frac{i (1-B_{\phi})}{p^2(1-B_{\phi})-(M^2_0+A_{\phi})+i\epsilon}
=\frac{i}{p^2-M^2+i\epsilon},
\end{eqnarray}
where $M^2=M_0^2(1+B)+A$ stands for the corresponding prediction for the
square of the mass at order $p^4$ \cite{comment3}
and $M^2_0$ at  order $p^2$.
In other words, the renormalized propagator is identical with the
{\em free} propagator in this approach.

It is easily shown using eqs.\ (\ref{wfrc}) to (\ref{rpgl}) that the
renormalized propagator in the presence of the off--shell terms is given by
\begin{equation}
\label{nprop}
i \Delta_R(p)= \frac{i}{p^2-M^2+\frac{16\beta_1}{F^2}(p^2-M^2)^2+i\epsilon},
\end{equation}
i.e., the renormalized propagator has now a nontrivial $p^2$ dependence.

\section{Discussion and conclusion}

The results for the form factors may now be summarized as follows:
\begin{eqnarray}
\label{formfactors}
F_{\pi^+}(q^2,p_f^2,p_i^2)&=&1+q^2 \left( 2 \frac{L^r_9}{F^2}
+2A(q^2,M^2_{\pi},\mu^2) +A(q^2,M^2_{K},\mu^2)\right) \nonumber \\
& & +\frac{16 \beta_1}{F^2} (p^2_f+p^2_i-2M^2_{\pi}),\nonumber \\
G_{\pi^+}(q^2,p_f^2,p_i^2)&=&(p_i^2-p_f^2)\left(2 \frac{L^r_9}{F^2}+
2A(q^2,M^2_{\pi},\mu^2) + A(q^2,M^2_{K},\mu^2)\right), \nonumber \\
F_{K^+}(q^2,p_f^2,p_i^2)&=&1+q^2\left(2 \frac{L^r_9}{F^2}
+2A(q^2,M^2_{K},\mu^2) +A(q^2,M^2_{\pi},\mu^2)\right) \nonumber \\
& & +\frac{16 \beta_1}{F^2} (p^2_f+p^2_i-2M^2_{K}),\nonumber \\
G_{K^+}(q^2,p_f^2,p_i^2)&=&(p_i^2-p_f^2)\left(2 \frac{L^r_9}{F^2}+
2A(q^2,M^2_{K},\mu^2) + A(q^2,M^2_{\pi},\mu^2)\right), \nonumber \\
F_{K^0}(q^2,p_f^2,p_i^2)&=&q^2(A(q^2,M^2_{K},\mu^2)-A(q^2,M^2_{\pi},\mu^2)),
\nonumber \\
G_{K^0}(q^2,p_f^2,p_i^2)&=&(p_i^2-p_f^2)(A(q^2,M^2_{K},\mu^2)
-A(q^2,M^2_{\pi},\mu^2)),
\end{eqnarray}

The function $A(q^2,M^2,\mu^2)$ of eq.\ (\ref{afunction}) is
scale--dependent through the term $\ln(M^2/\mu^2)$.
This scale dependence disappears from the form
factors, either because they explicitly contain the difference of
two such terms, as is the case for the $K^0$ form factors,
or because of a compensation with the scale dependence of the renormalized
parameter $L^r_9$, as is the case for the charged form
factors.
The parameters $L_9$ and $L_9^r$ are related through\footnote{For the
connection between  the coefficients $L_i$ of the
Gasser and Leutwyler lagrangian and the renormalized coefficients $L_i^r$
see eq.\ (7.25) of ref.\ \cite{Gasser2} and appendix B.2 of
ref.\ \cite{Donoghue}. We use the convention of ref.\ \cite{Donoghue}.}
$L_9=L_9^r+R/128 \pi^2$.

For $p_i^2=p_f^2=M^2$ the above expressions reduce to the on--shell
results of Gasser and Leutwyler \cite{Gasser3}.
For the charged pion and kaon the mean square radii are dominated by the
phenomenological parameter $L^r_9$ with rather moderate modifications from
the loop diagram of fig.\ 3, once renormalization is taken
into account.
The description of the radii is of reasonable quality \cite{Gasser3}.
The $K^0$ form factor is entirely given by loop diagrams.
The reason is, that the "minimal substitution type" coupling of
the covariant derivatives does not allow for a direct coupling
to neutral particles.

The terms proportional to $\beta_1$ in $F_{\pi^+}$ and $F_{K^+}$
are genuine off--shell terms.
Chiral perturbation theory makes definitive predictions for the off--shell
behavior in the following sense.
Regardless of the specific value of the constant $\beta_1$, it predicts
that the form factors of the charged pion and of the charged kaon
go off--shell {\em in the same way}, i.e., the slope at the real
photon point is in both cases given by the same $\beta_1$.
This is clearly a prediction of chiral symmetry, as otherwise there
is no compelling reason why the kaon and pion form factors should
be related.
Furthermore, chiral symmetry predicts in next--to--leading order
that the off--shell slope of the $K^0$ form factor vanishes.
Once again, this is a nontrivial result since gauge invariance
alone does not impose unique restrictions on the off--shell
behavior.

The contributions proportional to $\beta_2$ (see eqs.\ (\ref{fd4}) and
(\ref{ds})) have disappeared from the final result after wave function
renormalization.
This had to be the case, as otherwise the on--shell form factors would
not have reduced to the charge at $q^2=0$ (cf.\ eq.\ (\ref{fd4})).

The second off--shell manifestation is the appearance of the form factor
$G$.
Due to gauge invariance the form of $G$ is completely given by $F$
(see appendix A for a discussion of this point).
Here, we calculated $G$ independently and used the Ward--Takahashi
identity as a check of our calculation.
Using the results for the form factors of eq.\ (\ref{formfactors}),
the expression for the renormalized propagator of eq.\ (\ref{nprop})
and the Ward--Takahashi identity of eq.\ (\ref{rwti4}) or
(\ref{rwti5}) (see appendix A), respectively, it is straightforward to
verify that our calculation satisfies this important constraint.

A possible way to obtain information about the parameter $\beta_1$
is a comparison with effective lagrangians derived from a
quark model approach to $QCD$.
Such methods typically generate structures of the type of
eq.\ (\ref{addstr}) and the lowest--order equation of motion is commonly
used to bring it into the form of the standard Gasser and Leutwyler
lagrangian.
We consider the results of refs.\ \cite{Balog,Ebert,Espriu}
and bring them into the form of eqs.\ (\ref{l4gl}) plus (\ref{los}), i.e.,
we do not make use of the equation of motion.
One then finds $\beta_1=N_C/384 \pi^2=0.79\times10^{-3}$ .

In fig.\ 8 we present the on--shell form factor $F$ of the pion
together with the half--off--shell kinematics
$p_i^2=(1/2,-1/2)M^2_{\rho}$ and $p_f^2=M^2_{\pi}$.
It is commonly argued that the $\rho$ resonance is measure for the
convergence radius of the derivative expansion.
We restrict our considerations to values of
$|p^2|/M^2_{\rho}\leq 1/2$, where $p^2$ stands for $q^2,p_i^2,p_f^2$.
The curves are obtained with $\beta_1=0.79\times 10^{-3},$
and $L_9^r(M^2_{\eta})=7.4 \times 10^{-3}$.
The effect of the off--shell contribution is to shift the on--shell
curve, where the shift is proportional to the parameter $\beta_1$
and to the amount by which the momentum is off--shell.

Let us for example consider the t--channel kinematics in threshold pion
photo-- and electroproduction from the nucleon.
The photon momentum transfer squares of $q^2=(0,-0.05,-0.1)GeV^2$,
would correspond to $p_i^2=(-0.016,-0.060,-0.10) GeV^2
=(-0.027,-0.10,-0.18)M^2_{\rho}$, respectively.

The $G$ form factor is shown in fig.\ 9 with the same kinematics as in
fig.\ 8.
In fact, it shows little $q^2$ dependence, which can be understood
with the help of eq.\ (\ref{gitof}) of appendix A.
This is related to the fact that the prediction for $F$ is basically linear in
$q^2$.

In fig.\ 10 both the $F$ and $G$ form factors of the $K^0$ are
shown for $p_i^2=\pm M^2_{\rho}/2$.
Note the scale and the fact that they do not depend on $\beta_1$.
To the order we are considering only the form factor $G$ depends
on the off--shell kinematics.

Clearly, the predictions for the form factors are rather limited at
order $p^4$.
The reason is that the polarization vector of the photon and the extracted
momentum dependence in the vertex already provide two powers of $p$.
Thus to this order one can only predict either the slope with respect
to $q^2$ or to $p_i^2$ $(p_f^2)$.

It is straightforward to show that the results of eq.\ (\ref{formfactors})
reproduce those obtained with $PCAC$ and soft--pion
techniques\footnote{We would like to thank Barry R. Holstein for
drawing our attention to this point.} (see e.g.\ ref.\ \cite{Donoghue},
chpt.\ IV-5). In this context one has to realize that our predictions
are for the irreducible, renormalized Green's function
(see eq.\ (\ref{irtpgf})) whereas the reduction formula
used in the application of soft--pion techniques generates
the reducible Green's function of eq.\ (\ref{rtpgfm}).
Taking the renormalized propagator of eq.\ (\ref{nprop}) into account
our predictions are in agreement with the soft--pion results.

The lagrangians of the standard Gasser and Leutwyler approach and the
above off--shell extension are completely equivalent with respect to
on--shell matrix elements.
They only yield different predictions if one considers Greens's functions
off--shell.
The use of the classical equation of motion to eliminate additional structures
can be interpreted as a generalized field transformation of the interpolating
field \cite{Leutwyler,Georgi}.
The equivalence theorem \cite{Haag,Kamefuchi,Coleman} then guarantees
identical S--matrix elements, even though individual building blocks,
e.g.\ off--shell Green's functions, may be different.
That this is in fact the case in the above example will be discussed
elsewhere using Compton scattering \cite{Fearing}.
Thus, the use of the equation of motion is a means to bring the effective
lagrangian into its most efficient form, i.e., the one with the least free
parameters.

In conclusion, we have considered the most general electromagnetic vertex
of pions and kaons compatible with chiral symmetry at order $p^4$.
The off--shell extension allows for two additional structures in the
effective lagrangian at order $p^4$.
After renormalization only one additional term beyond the standard
Gasser and Leutwyler lagrangian contributes to the off--shell vertex.
Chiral symmetry predicts that the form factors $F$ of the charged pions
and kaons go off--shell in the same way.
In both cases the off--shell slope at the real photon point is
proportional to a new parameter $\beta_1$.
Furthermore, at order $p^4$ chiral symmetry makes a unique prediction
how the neutral kaon form factors go off--shell, as the corresponding result
is entirely due to one--loop diagrams without tree--level contribution.

\section{Acknowledgements}

This work was supported in part by a grant from the Natural Sciences and
Engineering Research Council of Canada. The authors would like to thank
Barry R.\ Holstein, Justus H.\ Koch and Dong--Pil Min for useful comments.

\begin{appendix}

\section{The Ward--Takahashi identity}

Powerful constraints on the off--shell form factors result from the
Ward--Takahashi identity \cite{Ward,Takahashi}.
The standard equal--time commutation relations between the charge density
and the field operators resulting from $U(1)$ gauge invariance,
\begin{eqnarray}
\label{comrel}
[J^0(x),\pi^-(y)] \delta (x^0-y^0) & = & \delta^4(x-y) \pi^-(y), \nonumber \\
{[}J^0(x),\pi^+(y)]  \delta (x^0-y^0) & = & -\delta^4(x-y) \pi^+(y),
\end{eqnarray}
and current conservation, $\partial_{\mu} J^{\mu}(x)=0$, are the main
ingredients for obtaining the Ward--Takahashi identity \cite{wtder},
\begin{equation}
\label{rwti3}
q_{\mu} \Gamma^{\mu,irr}_R(p_f,p_i) =
\Delta_R^{-1}(p_f)-\Delta_R^{-1}(p_i).
\end{equation}
Inserting the parameterization of the irreducible vertex, eq.\ (\ref{par1}),
into the Ward--Takahashi identity, eq.\ (\ref{rwti3}), the constraint
on the form factors $F$ and $G$ reads
\begin{equation}
\label{rwti4}
(p_f^2-p_i^2) F(q^2,p_f^2,p_i^2)+q^2 G(q^2,p_f^2,p_i^2)
= \Delta^{-1}_R(p_f)-\Delta^{-1}_R(p_i).
\end{equation}
For neutral particles, e.g.\ the $K^0$ and $\bar{K}^0$, the situation
is even simpler, since the field operators commute with the charge
density and thus the right--hand side of eq.\ (\ref{rwti3}) vanishes.
The Ward--Takahashi identity then reads
\begin{equation}
\label{rwti5}
(p_f^2-p_i^2) F(q^2,p_f^2,p_i^2)+q^2 G(q^2,p_f^2,p_i^2)=0.
\end{equation}

In the following, we discuss some of the consequences of the Ward--Takahashi
identity, eq.\ (\ref{rwti4}), for the form factors $F$ and $G$ of charged
particles.
We first consider the case of real photons, $q^2=0$.
Using $\Delta_R^{-1}(p)=0$ for $p^2=M^2$,
the Ward--Takahashi identity for the half--off--shell
situation\footnote{Here one assumes that $G(q^2,p_f^2,p_i^2)$
is not singular for $q^2=0$, which is certainly reasonable as long as there
are no strongly interacting zero--mass particles in the theory.}
reads
\begin{equation}
\label{cond1}
\Delta^{-1}_R(p)=(p^2-M^2) F(0,p^2,M^2)=(p^2-M^2) F(0,M^2,p^2).
\end{equation}
Using the standard parameterization \cite{Cheng,Itzykson}
\begin{equation}
\label{invprop}
\Delta^{-1}_R(p)=p^2-M^2-\Pi(p^2)+i\epsilon,
\end{equation}
with the normalization conditions
\begin{equation}
\label{normcond}
\Pi(M^2)=\Pi'(M^2)=0,
\end{equation}
one finds the standard interpretation of $F(0,M^2,M^2)=1$ as the charge of
the pion \cite{Barton} by multiplying eq.\ (\ref{cond1}) by the inverse of
$p^2-M^2$ and taking the limit $p^2 \rightarrow M^2$ .
However, since $((p^2-M^2)\Delta_R(p))^{-1}\neq 1$ in general,
the extension of such an interpretation to the half--off--shell
case is in general not possible \cite{Naus1}.

The complete off--shell case at the real photon point can be expressed
in terms of the half--off--shell case by means of eqs.\ (\ref{rwti4}),
(\ref{cond1}),
\begin{equation}
\label{cond2}
F(0,p_f^2,p_i^2)=
\frac{(p_f^2-M^2)F(0,p_f^2,M^2)-(p_i^2-M^2)F(0,M^2,p_i^2)}{p_f^2-p_i^2}.
\end{equation}
Here it is worthwhile to notice that, using eq.\ (\ref{cond1}), we obtain
the relation $F(0,p_f^2,p_i^2)=F(0,p_i^2,p_f^2)$ (see eq.\ (\ref{tri}))
even without time reversal symmetry, however, only at the real photon
point.

We now turn to the case of virtual photons, $q^2\neq 0$.
Using eqs.\ (\ref{rwti4}), (\ref{cond1}) and (\ref{cond2})
we can express $G$ in terms of $F$
\begin{equation}
\label{gitof}
G(q^2,p_f^2,p_i^2)=
\frac{(p^2_f-p^2_i)\left(F(0,p^2_f,p^2_i)-F(q^2,p^2_f,p^2_i)\right)}{q^2}.
\end{equation}
Considering the half--off--shell limit of eq.\ (\ref{gitof}),
\begin{equation}
\label{ghos}
G(q^2,p^2,M^2)=-G(q^2,M^2,p^2)=
\frac{(p^2-M^2)\left(F(0,p^2,M^2)-F(q^2,p^2,M^2)\right)}{q^2}.
\end{equation}
one finds after differentiating eq.\ (\ref{ghos}) with respect to
$p^2$ and taking the limit $p^2\rightarrow M^2$ a relation between
the on--shell form factor $F(q^2,M^2,M^2)$ and the way the form factor
$G$ goes off--shell
\begin{equation}
\label{fg}
F(q^2,M^2,M^2)=1+q^2 \frac{\partial G(q^2,M^2,M^2)}{\partial p_i^2}
=1-q^2 \frac{\partial G(q^2,M^2,M^2)}{\partial p_f^2}.
\end{equation}

The contents of the Ward--Takahashi identity for spinless, charged particles
can be summarized by saying that the knowledge of $F(q^2,p_f^2,p^2_i)$
completely determines the propagator (self energy) of the
particle (see eq.\ (\ref{cond1}))
as well as the form factor $G(q^2,p_f^2,p^2_i)$ (see eq.\ (\ref{gitof})).

\section{Alternative representation of the electromagnetic vertex}

In this section we will relate the parameterization of the irreducible
vertex of eq.\ (\ref{par1})
to other conventions used in the literature\footnote{Here we will
denote the renormalized, irreducible vertex simply by $\Gamma^{\mu}$.}.
Following refs. \cite{Nishijima,Ohta}, $\Gamma^{\mu}$ can also be written as
\begin{equation}
\label{vn}
\Gamma^{\mu}(p_f,p_i)=P^{\mu} A(q^2,p_f^2,p_i^2)
+\left(P^{\mu} q^2-q^{\mu}P\cdot q\right)B(q^2,p_f^2,p_i^2),
\end{equation}
where $P^{\mu}=p_f^{\mu}+p_i^{\mu}$ and $q^{\mu}=p_f^{\mu}-p_i^{\mu}$.
Due to time--reversal invariance the functions $A(q^2,p_f^2,p_i^2)$ and
$B(q^2,p_f^2,p_i^2)$ are symmetric in $p_i^2$ and $p_f^2$
(see eq.\ (\ref{tri})).
In the above notation the second operator structure is separately gauge
invariant, i.e., vanishes when contracted with $q_{\mu}$.
Thus the Ward--Takahashi identity only involves the function
$A(q^2,p_f^2,p_i^2)$, namely
\begin{equation}
\label{wtn}
q_{\mu}\Gamma^{\mu}(p_f,p_i)=
\left(p_f^2-p_i^2\right)A(q^2,p_f^2,p_i^2)=
\Delta_R^{-1}(p_f)-\Delta_R^{-1}(p_i).
\end{equation}
As the right--hand side of eq.\ (\ref{wtn}) involves functions of either
$p_f^2$ or $p_i^2$, only, taking the partial derivative with respect to
$q^2$ shows that the function $A$ does not actually depend on $q^2$.
Clearly, in the above representation, a complete determination of the
vertex function, $\Gamma^{\mu}$, requires the knowledge of the full propagator
$\Delta_R(p)$ and of the function $B(q^2,p_f^2,p_i^2)$.
It is straightforward to relate the functions $A$ and $B$ to the functions
$F$ and $G$ of eq.\ (\ref{par1}),
\begin{eqnarray}
\label{relabfg}
A(q^2,p_f^2,p_i^2) & = & F(q^2,p_f^2,p_i^2)
+\frac{q^2}{P\cdot q} G(q^2,p_f^2,p_i^2)=
F(0,p^2_f,p^2_i),\nonumber \\
B(q^2,p_f^2,p_i^2) & = & - \frac{G(q^2,p_f^2,p_i^2)}{P \cdot q}=
\frac{F(q^2,p_f^2,p_i^2)-F(0,p^2_f,p^2_i)}{q^2}.
\end{eqnarray}
Another closely related representation is, e.g., used in ref.\ \cite{Gross},
\begin{equation}
\label{vg}
\Gamma^{\mu}(p_f,p_i)=
\left(P^{\mu} -q^{\mu} \frac{P\cdot q}{q^2}\right) {\cal A}(q^2,p_f^2,p_i^2)
+P^{\mu} {\cal B}(q^2,p_f^2,p_i^2),
\end{equation}
and comparing with eq.\ (\ref{vn}) one finds
\begin{eqnarray}
\label{aabb}
{\cal A}(q^2,p_f^2,p_i^2) & = & q^2 B(q^2,p_f^2,p_i^2), \nonumber \\
{\cal B}(q^2,p_f^2,p_i^2) & = & A(q^2,p_f^2,p_i^2).
\end{eqnarray}
\end{appendix}

\newpage
{\Large \bf Figure captions}
\vspace{1cm}

Fig.\ 1. Tree--level diagram: The vertex is derived from ${\cal L}_2$, denoted
by 2 in the interaction blob.

Fig.\ 2. 1--vertex loop diagram: The vertex is derived from ${\cal L}_2$,
denoted by 2 in the interaction blob.

Fig.\ 3. 2--vertex loop diagram: The vertices are derived from
${\cal L}_2$, denoted by 2 in the interaction blobs.
Note that only charged pions and kaons appear in the loop.

Fig.\ 4. Tree--level diagram: The vertex is derived from
${\cal L}_4$, denoted by 4 in the interaction blob.

Fig.\ 5. Typical 2--vertex loop diagram: The vertices are derived from
${\cal L}_2$, denoted by 2 in the interaction blobs.

Fig.\ 6. 1--vertex loop contribution to the self energy: The vertex is
derived from
${\cal L}_2$, denoted by 2 in the interaction blob.

Fig.\ 7. Tree--level contribution to the self energy: The vertex is
derived from
${\cal L}_4$, denoted by 4 in the interaction blob.

Fig.\ 8. Pion form factor $F$: The solid line corresponds to
the on--shell form factor. The other two curves are obtained with
$p_f^2=M^2_{\pi}$ and $p_i^2=M^2_{\rho}/2$ (dashed line) and
$p_i^2=-M^2_{\rho}/2$ (dashed--dotted line), respectively.
The curves are obtained with $\beta_1=0.79\times 10^{-3}$
and $L^r_9(M^2_{\eta})=7.4\times 10^{-3}$.

Fig.\ 9. Off--shell pion form factor $G$: The two curves correspond
to $p_f^2=M^2_{\pi}$ and $p_i^2=M^2_{\rho}/2$ (dashed line) and
$p_i^2=-M^2_{\rho}/2$ (dashed--dotted line), respectively.
Note that $G$ vanishes on--shell.
The curves are obtained with $\beta_1=0.79\times 10^{-3}$
and $L^r_9(M^2_{\eta})=7.4\times 10^{-3}$.

Fig.\ 10. Off--shell form factors $F$ and $G$ of the $K^0$:
The solid curve corresponds to the form factor $F$ which
is independent of off--shell kinematics at order $p^4$.
The other two curves represent the $G$ form factor for
$p_f^2=M^2_{\pi}$ and $p_i^2=M^2_{\rho}/2$ (dashed line) and
$p_i^2=- M^2_{\rho}/2$ (dashed--dotted line), respectively.
The result is independent of $\beta_1$ (and $L^r_9$) and thus unique.

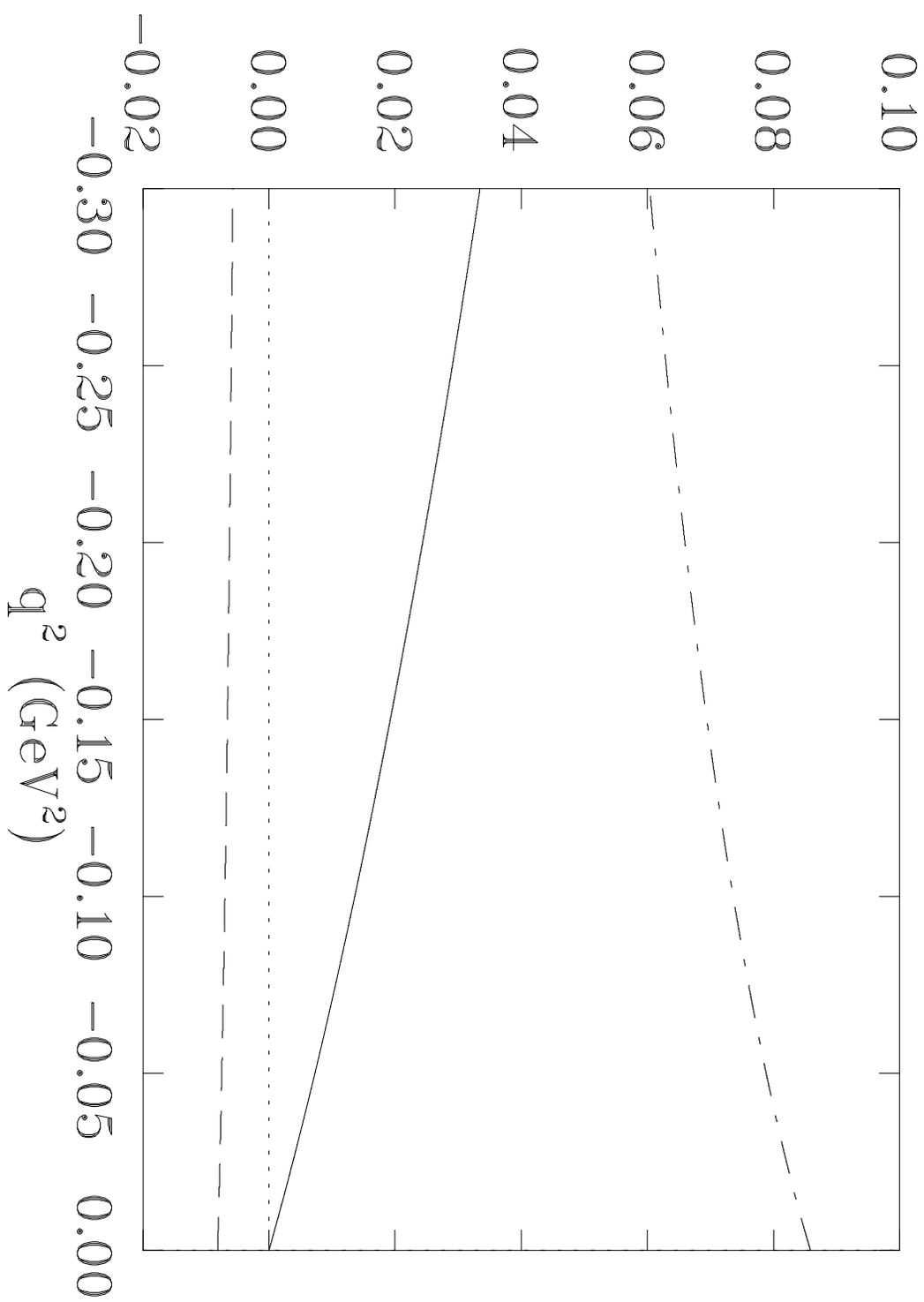
\end{document}